\documentclass[prb,twocolumn,showpacs,preprintnumbers]{revtex4}

\usepackage{amsmath,amssymb,amsfonts}
\usepackage{dcolumn}
\usepackage{bm}
\usepackage{graphicx}

\begin{document}

\title{Metal--Insulator Transitions in the Falicov--Kimball Model with Disorder}
\author{Krzysztof Byczuk}
\affiliation{Institute of Theoretical Physics, Warsaw University,
  ul. Ho\.za 69, PL-00-681 Warszawa, Poland}

\date{\today}

\begin{abstract}
The ground state phase diagrams of the  Falicov--Kimball model with local
disorder is
derived within  the dynamical mean--field theory  and using the geometrically
averaged (''typical'') local density of states. Correlated metal,
Mott insulator and Anderson insulator
phases are identified. The metal--insulator transitions are  found to be continuous.
The interaction and disorder compete with each other stabilizing the metallic
phase against occurring one of the insulators. 
The Mott and Anderson insulators are found to be
 continuously connected.

\end{abstract}
\maketitle




\section{Introduction}

Motion of quantum particles can be suppressed or even destroyed by Coulomb 
interactions and disorder, which are the driving forces behind  a 
metal--insulator transition (MIT). 
The Mott--Hubbard MIT is caused by Coulomb correlations in 
the pure system, i.e. without disorder.\cite{Mott90} 
The Anderson MIT, also refereed to the Anderson localization, is
 due to coherent backscattering from randomly distributed impurities 
in a system without interaction.\cite{Anderson58}
The properties of real materials are strongly influenced by both 
interaction and  randomness.\cite{Kravchenko}
It is therefore a challenge to investigate quantum models  
where both correlations and disorder are simultaneously 
present.\cite{Lee85,Belitz94,ma,byczuk04}

The Mott--Hubbard MIT is characterized by opening a gap in the density of states 
at the Fermi level.\cite{hubbard} 
At the Anderson localization the character of the 
spectrum at the Fermi level changes from continuous one 
to dense pure point one.\cite{point} 
It is plausible that both MITs could be detected 
by knowing a single quantity, namely, a local density of states (LDOS).
Although the LDOS is not an order parameter associated 
with a symmetry breaking phase transition,\cite{Belitz94} it
discriminates between a metal and an insulator, which is driven by correlations and 
disorder. 

In a disordered system the LDOS depends on particular realization of the disorder. 
Then the entire probability distribution function of the LDOS is required
to know,\cite{Mirlin94} which  is a very demanding task. 
Instead one could use certain moments of the LDOS.
This however is insufficient because the 
arithmetically averaged LDOS (first moment) stays finite at the Anderson
MIT.\cite{Lloyd+Thouless}   
It was already pointed out by Anderson\cite{Anderson58}
 that the ``typical'' values of random quantities,
 which are  mathematically given by the most probable 
values of the probability distribution functions,\cite{definition} 
should be used  to describe localization. 
The \emph{geometric} mean \cite{lognormal,geometrical} gives an  
approximation of the most probable (``typical'')
value of the LDOS and vanishes at a critical strength of the disorder, hence
providing  an explicit criterion for Anderson 
localization.\cite{Anderson58,Dobrosavljevic97,Dobrosavljevic03,Schubert03}

Theoretical descriptions 
of the MIT has  to be non--perturbative if no long--range order exists 
on either side of the transition. 
In such a case 
there is no obvious order parameter and no Landau type functional available. 
A non--perturbative framework for investigation of the Mott--Hubbard MIT
in  lattice electrons with a local interaction and disorder  
is given  by dynamical mean--field theory (DMFT).\cite{metzner89,georges96,pruschke95,vollhardt93}
If in this approach the effect of local disorder is taken into account
through the arithmetic mean of the LDOS \cite{ulmke95} one obtains, in the
absence of interactions, the well known coherent potential approximation (CPA),\cite{vlaming92}
 which does not describe the physics of  Anderson
localization. To overcome this deficiency Dobrosavljevi\'{c} and 
Kotliar\cite{Dobrosavljevic97} formulated a variant of the DMFT where the
geometrically averaged LDOS is computed from the solutions of the
self--consistent stochastic DMFT equations.
Subsequently, Dobrosavljevi\'{c} \emph{et al.} \cite{Dobrosavljevic03}
incorporated the geometrically averaged LDOS into the self--consistency cycle and
thereby derived  a mean--field theory of Anderson localization which
reproduces many of the expected features of the disorder--driven
MIT for non--interacting electrons.
This scheme  uses only one--particle quantities and is therefore
easily incorporated into the DMFT for disordered electrons in the presence of
phonons,\cite{fehske} or Coulomb correlations.\cite{byczuk04}
In particular, the non--magnetic ground state phase diagram of the Anderson--Hubbard model at 
half--filling was derived.\cite{byczuk04}

In this paper we investigate the Falicov--Kimball model\cite{falicov} with a local disorder. 
The pure Falicov--Kimball model describes two species of particles,
mobile and immobile, which interact with each other when both 
are on the same lattice site.\cite{falicov, freericks} 
The Falicov--Kimball model captures some aspects of the Mott--Hubbard MIT, i.e. upon increasing 
the interaction the LDOS for mobile particles splits into two subbands opening 
a correlation gap at the Fermi level if $n_e+n_f=1$, where $n_e$ ($n_f$) is the density of 
mobile (immobile) particles.\cite{dongen,freericks,lemanski,freericks04}
Here we introduce the \emph{Anderson--Falicov--Kimball model}
 where the mobile particles are disturbed by a local 
random potential. 
Our aim is to obtain a  phase diagram of such a model and to 
identify MITs driven by correlations
and disorder.
We find a subtle competition between interaction and disorder yielding stabilization of  metalicity 
in the  Anderson--Falicov--Kimball model. 
The model is solved within the DMFT framework combined with geometric averaging of the LDOS. 
The results are compared with those obtained within the DMFT with  arithmetic averages. 
Only geometric averaging yields the Anderson transition. 

In Section II we define the Anderson--Falicov--Kimball model and present the
DMFT equations which provide the solution of this model. 
In Section III and IV  numerical results
concerning the ground state phase diagram are shown and discussed in
details. The analytical approach to the MIT in the band center is developed in 
Section V.
Section VI presents our conclusions and final remarks.

\section{Anderson--Falicov--Kimball Model}

\subsection{The Model}

The Anderson--Falicov--Kimball model is defined by the following Hamiltonian 
\begin{equation}
H=\sum_{ij}t_{ij}c^{\dagger} _i c_j + \sum_i \epsilon_i c^{\dagger} _i c_i
+ U\sum_i f^{\dagger} _i f_i c_i^{\dagger}c_i ,
\label{fk}
\end{equation}
where $c_i^{\dagger}$ ($f_i^{\dagger}$) and $c_i$ ($f_i$) are fermionic creation and annihilation
operators for \emph{mobile}  (\emph{immobile}) particles
at a lattice site $i$.
$t_{ij}$ is a hoping amplitude for mobile particles between sites $i$ and $j$,
and $U$ is the local interaction energy between mobile and  immobile particles
occupying the same  site.
The ionic energy $\epsilon_i$ is a random, independent variable in our problem,
describing the local disorder disturbing a motion of mobile particles.
We have assumed that only mobile particles are subjected to the structural disorder.
This assumption could be relaxed in further generalizations of the model.

The position of the immobile particles on a lattice 
 is random if there is no long-range order. Therefore, we 
assume that   the occupation number $f^{\dagger} _i f_i $ on the  $i$-th site   is 
equal to one with probability $w$ ($0\leq w \leq 1$) and zero with probability $1-w$. 
The presence of the randomly distributed immobile particles introduces additional disorder apart of that
given by the $\epsilon_i$-term in the Hamiltonian (\ref{fk}). 
However, the $U$-term in the Hamiltonian (\ref{fk}) must be treated differently from the $\epsilon_i$-term. 
The $U$-term is operator valued  in the immobile particle Fock subspace and one has to take 
the quantum mechanical average over a given quantum state of the $f$-particles. 
In contrast, the $\epsilon_i$-term does not depend on the $f$ operators and one has to average the 
quantum mechanical expectation values over 
 different realizations of $\epsilon_i$ or one has to study the whole statistics of an interesting 
quantum mechanical expectation value. 
Whereas, the  $U$-term does not change the extended states into the localized ones, the 
$\epsilon_i$-term can lead to such a change and thereby to the Anderson localization. 

In this paper we use the canonical description, where the number of the immobile and mobile particles
are independent of each other and fixed.
In the pure Falicov--Kimball model increasing the interaction leads to opening 
a correlation (Mott) gap in the spectrum at $U=U_c$.\cite{dongen,freericks,lemanski,freericks04} 
If $n_e+n_f=1$ the Fermi energy for mobile particles is inside of this correlation gap, which means that 
the ground state is incompressible.
On the Bethe lattice with the band--width $W$  the critical interaction
obtained within the DMFT is
$U_c=W/2$ for $n_e=n_f=1/2$.\cite{dongen,freericks,lemanski,freericks04} 
How the disorder changes this gap and how the localized states enter into the mobile particle band are 
the subjects of the present study.
We neglect any long--range order, which might be achieved on a fully 
frustrated lattice.\cite{georges96}

\subsection{Dynamical Mean--Field Theory}

The Anderson--Falicov--Kimball model, where the interaction and  disorder are local, 
is solved within the DMFT.\cite{brand,georges96,pruschke95,vollhardt93}
Introducing a single and double-particle (Zubarev) Green functions\cite{zubariev}
$G_{ij}(\omega)=\langle \langle c_i|c^{\dagger}_j \rangle \rangle_{\omega}$ and 
$\Gamma_{ij}(\omega)=\langle \langle f^{\dagger} _i f_i  c_i|c^{\dagger}_j \rangle \rangle_{\omega}$, 
respectively, we
derive  the equations of motion
\begin{subequations}
    \begin{align}
  &   (\omega+\mu-\epsilon_k)G_{kl}(\omega) - \sum_j
      t_{kj}G_{jl}(\omega)=\delta_{kl}+U\Gamma_{kl}(\omega) ,
  \\
   &   (\omega+\mu-\epsilon_k - U)\Gamma_{kl}(\omega) = \langle f^{\dagger} _k f_k \rangle
      \delta_{kl} + \sum_j t_{kj} \Gamma_{jl}(\omega),
    \end{align}
    \label{eom}
  \end{subequations}
\noindent where we used that a number of the immobile particles is conserved being zero or one, and hence 
$f^{\dagger} _i f_if^{\dagger} _i f_i  =f^{\dagger} _i f_i $.
The chemical potential $\mu$ is introduced
 only for the mobile subsystem and $\omega$ denotes the energy.
Defining the site--independent self--energy, according to DMFT scheme,\cite{brand,bulla_bis} 
\begin{equation}
\Sigma(\omega,\epsilon_i)\equiv U \frac{\Gamma_{ij}(\omega)}{G_{ij}(\omega)},
\end{equation}
which depends implicitly on $\epsilon_i$, the system of the equations
(\ref{eom}) can be solved yielding an explicite formula for the self--energy
\begin{equation}
\Sigma(\omega,\epsilon_i)=wU+\frac{w(1-w)U^2}{\omega+\mu-\epsilon_i-(1-w)U-\eta(\omega)}.
\label{sigma}
\end{equation}
We defined the averaged number  of localized particles per site 
$\langle f^{\dagger} _i f_i  \rangle=n_f=w$ 
 and we introduced the hybridization function
$\eta(\omega)$ which is a \emph{dynamical mean field} (\emph{molecular field}) describing
 the coupling of a selected lattice site with a rest of the system.
The local non--interacting Green function is 
$G^0_{ii}(\omega)=1/[\omega+\mu-\epsilon_i-\eta(\omega)]\equiv G^0(\omega,\epsilon_i)$.

Using  the self--energy $\Sigma(\omega,\epsilon_i)$ 
and the hybridization function $\eta(\omega)$ we obtain a local
($\epsilon_i$-dependent) Green function 
\begin{equation}
G_{ii}(\omega)=\frac{1}{\omega+\mu-\epsilon_i-\eta(\omega)-\Sigma(\omega,\epsilon_i)}\equiv 
G(\omega,\epsilon_i),
\label{green}
\end{equation}
and hence the $\epsilon_i$-dependent LDOS
\begin{equation}
A(\omega,\epsilon_i)=-\frac{1}{\pi} \rm{Im} G(\omega,\epsilon_i).
\label{ldoss}
\end{equation}
From the $\epsilon_i$-dependent LDOS (\ref{ldoss}) we obtain
either the geometrically averaged LDOS
\begin{equation}
A _{\mathrm{geom}}(\omega )=\exp \left[ \langle \ln A(\omega , \epsilon_i)\rangle_{\rm dis} \right]
\end{equation}
 or the arithmetically averaged LDOS
\begin{equation}
A _{\mathrm{arith}}(\omega )= \langle A(\omega,\epsilon_i)\rangle_{\rm dis}  ,
\end{equation} 
where  $\langle O(\epsilon_i) \rangle_{\rm dis} =\int d\epsilon
_{i}\mathcal{P}(\epsilon _{i})O(\epsilon _{i})$ is an arithmetic
mean of $O(\epsilon_i)$.\cite{comment2}
Here we used that $\epsilon_i$ are independent random variables 
characterized  by a probability distribution 
function $\mathcal{P}(\epsilon _{i})$.
The lattice, i.e. translationally invariant, Green function is given by the
corresponding Hilbert transform 
\begin{equation}
G(\omega )=\int d\omega
^{\prime }\frac{A _{\mathrm{\alpha}}(\omega ' )}{\omega -\omega
^{\prime }},
\end{equation} 
where the subscript $\mathrm{\alpha}$ stands for
either "$\mathrm{geom}$" or "$\mathrm{arith}$". 
The Dyson self--energy $\Sigma (\omega )$ is determined from the
$\mathbf{k}$-integrated Dyson equation $\Sigma(\omega )=\omega
-\eta (\omega )-1/G(\omega )$. 
The self--consistent DMFT equations are
closed through the Hilbert transform $ G(\omega )=\int d\epsilon
N_{0}(\epsilon )/\left[\omega -\epsilon -\Sigma (\omega )\right]
$, which relates the lattice Green function to
the self--energy; here $N_{0}(\epsilon )$ is the non--interacting
density of states.

The DMFT which uses $A_{\rm arith}$ is an exact approach in the limit of infinite dimension
 where quantum mechanical rescaling is imposed on hopping amplitudes.\cite{metzner89,vlaming92}
On the other hand, the mathematically 
rigorous limit for the DMFT with $A_{\rm geom}$  is not yet known.\cite{Dobrosavljevic03} 
Nevertheless, it is very promising single site theory which has the ability to describe at least 
some aspect of the Anderson localization, i.e. the localization due to random fluctuations of the 
wave function amplitude.\cite{economou} 
As a single site theory, it cannot capture interference effects due to fluctuations of a phase of the 
wave function and, thereby, weak localization aspects are not recovered.

The Anderson--Falicov--Kimball model (\ref{fk}) is solved for
a semi-elliptic density of states for the Bethe lattice, $N_{0}(\epsilon
)=4\sqrt{1-4(\epsilon/W) ^{2}}/(\pi W)$. 
Then $\eta (\omega )=W^2 G(\omega )/16$.
For a probability distribution function $ \mathcal{P}(\epsilon _{i})$ we assume a box model, i.e.
$\mathcal{P}(\epsilon _{i})=\Theta (\Delta /2-|\epsilon
_{i}|)/\Delta $, with $\Theta $ as the step function. The
parameter $\Delta $ is a measure of the disorder strength.
For numerical integrations  we use discrete values of $\epsilon_i$
selected according to the Gauss-Legendre algorithm.
The number of $\epsilon_i$ levels depends on $\Delta$
and is adjusted such to obtain  smooth density of states.
The chemical potential $\mu=U/2$, corresponding to half--filled conducting band (i.e., $n_e=1/2$), and
$w=1/2$ are assumed in this paper. $W=1$ sets the energy units.

\subsection{Criteria for Anderson and Mott MIT}

The arithmetically averaged LDOS
$A_{\mathrm{arith}}(\omega)$ at the energy $\omega$ in a band 
is always positive for non-interacting systems
with disorder.\cite{Lloyd+Thouless}
This quantity is non-critical for the Anderson localization. 
However it approaches zero when the  gap is opened at energy $\omega$ 
in the spectrum of the  correlated system. 
On the other hand $A_{\mathrm{geom}}(\omega)$ vanishes at the Anderson
localization. 
We therefore classify  the states at energy $\omega$  as
localized by disorder if 
$A_{\mathrm{geom}}(\omega)=0$ and $A_{\mathrm{arith}}(\omega)>0$. When both
$A_{\mathrm{geom}}(\omega)=0$ and $A_{\mathrm{arith}}(\omega)=0$ it means that the states
at energy $\omega$ are absent due to correlations.

\section{Spectral Phase Diagrams}

\begin{figure}[twb]
\centerline{\includegraphics [clip,width=8.5cm,angle=-0]{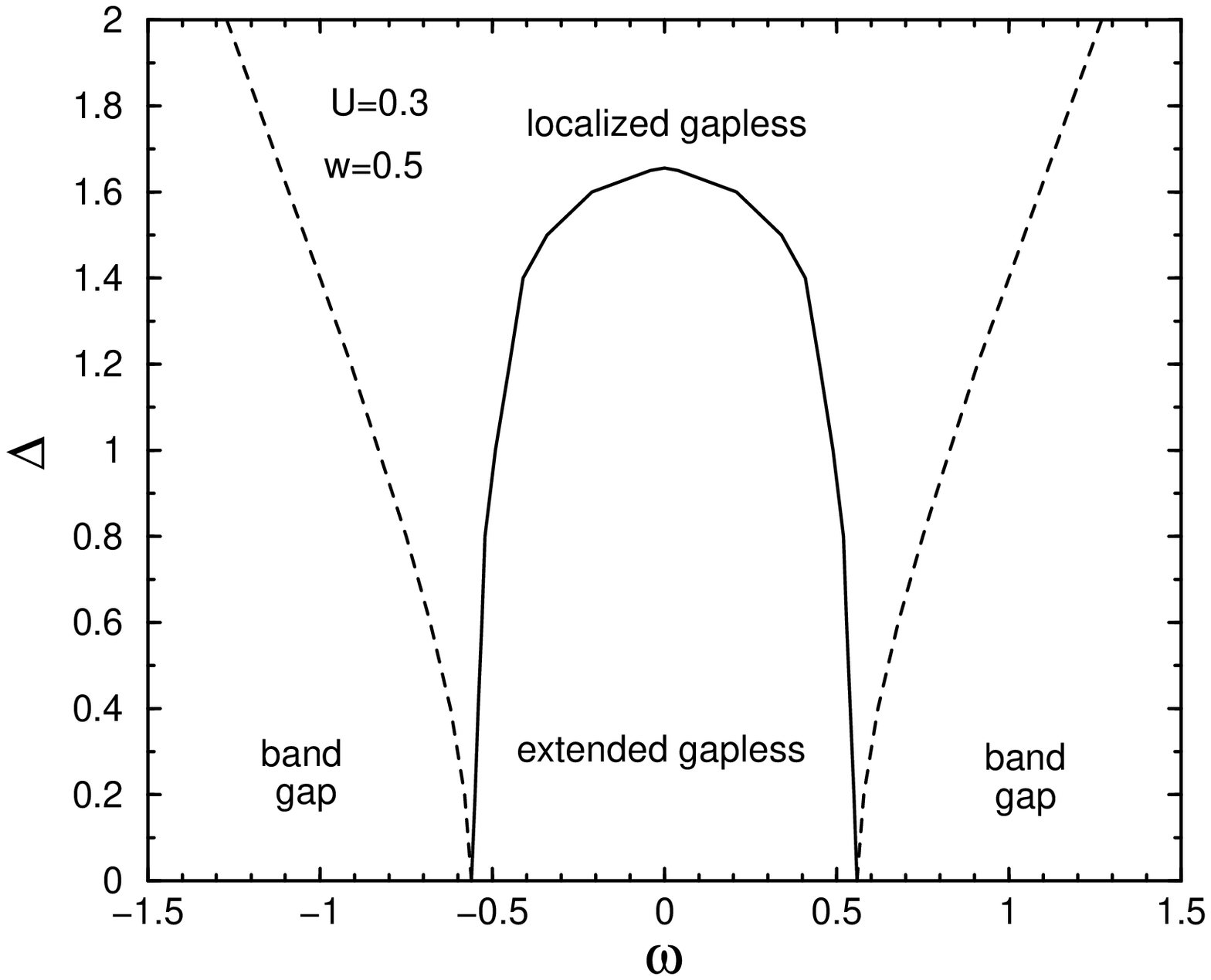}}
\nopagebreak
\centerline{\includegraphics [clip,width=8.5cm,angle=-0]{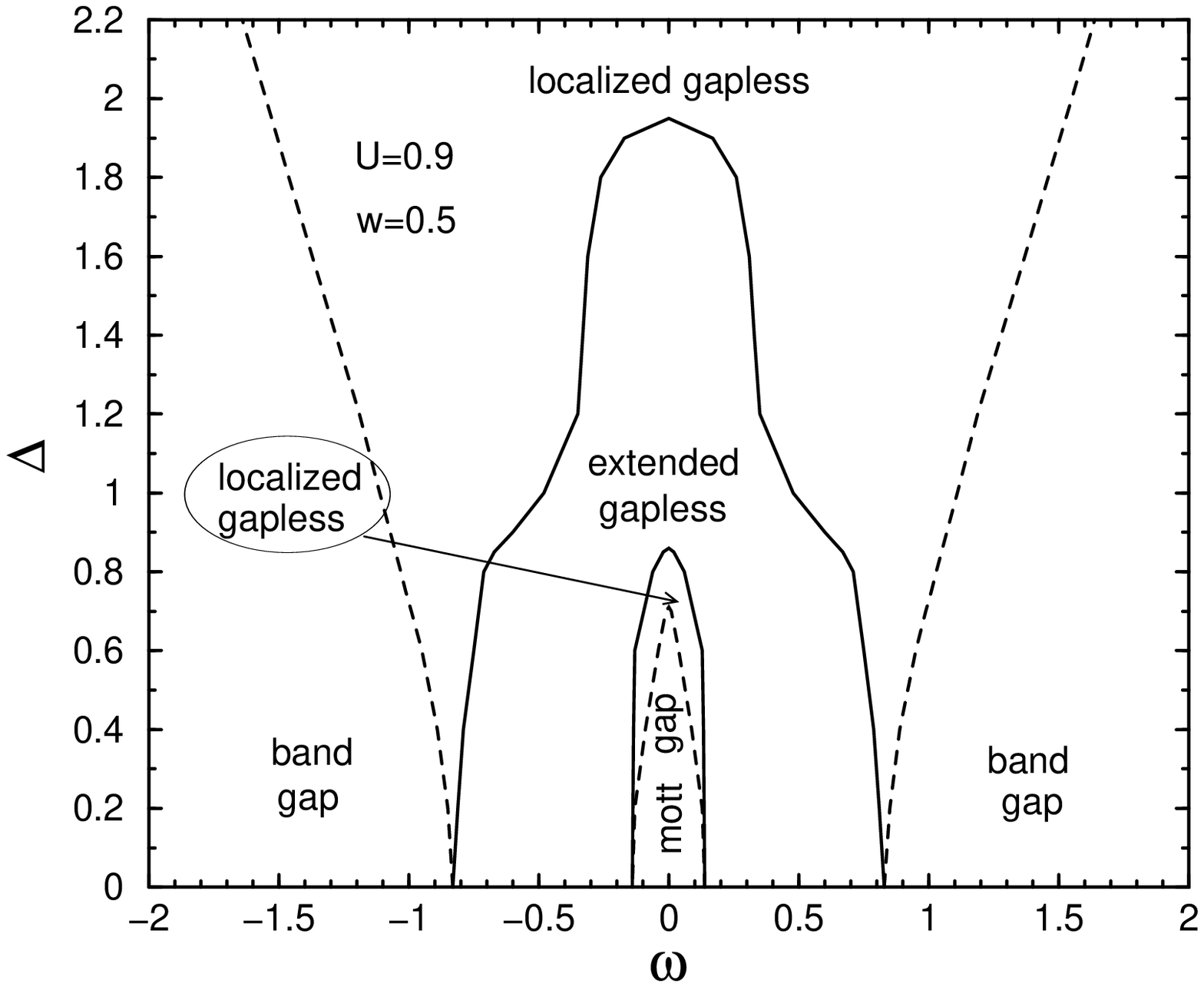}}
\nopagebreak
\centerline{\includegraphics [clip,width=8.5cm,angle=-0]{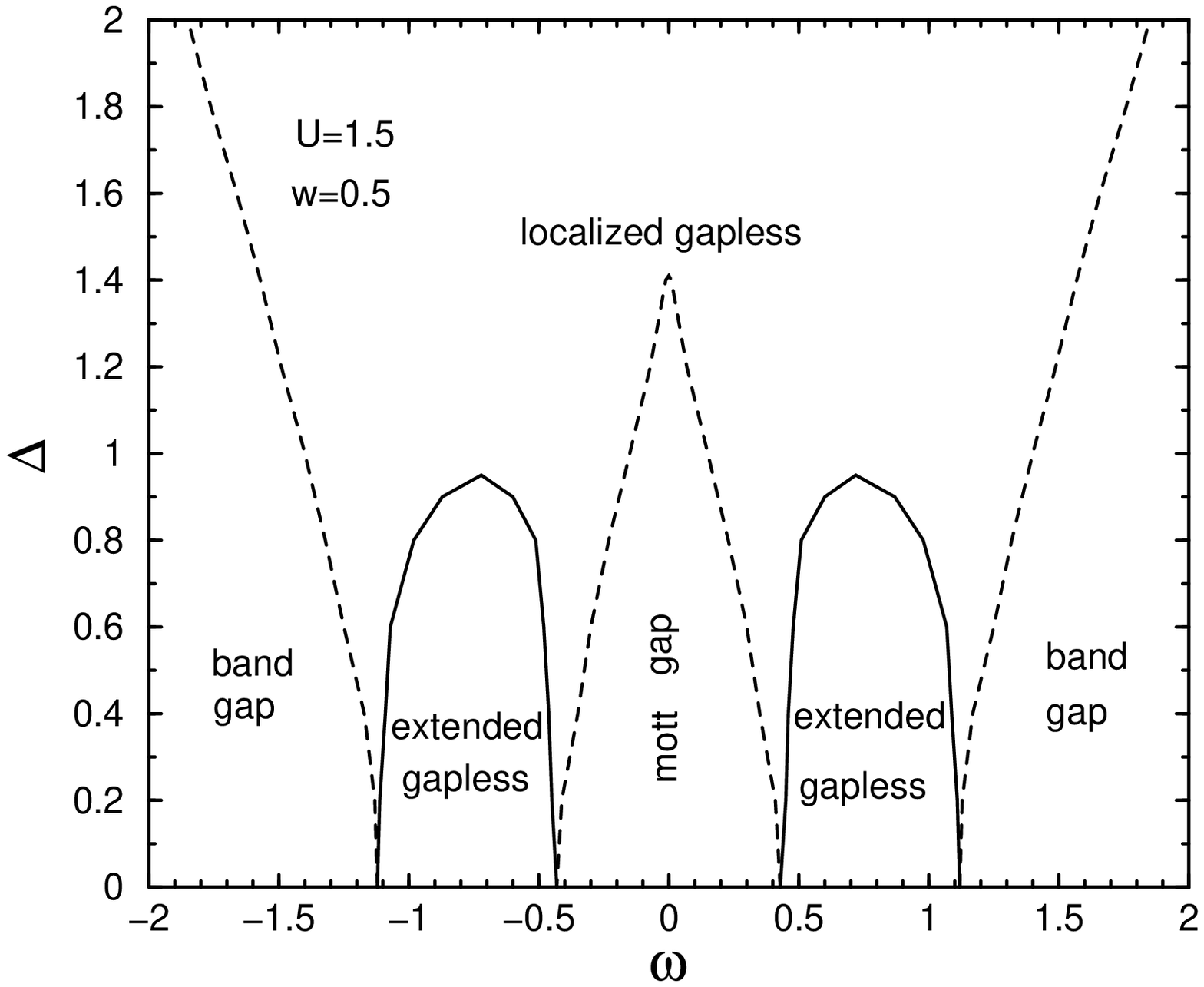}}
\nopagebreak
\caption{Spectral phase diagrams of the  ground states without a long--range
  orders for the Anderson--Falicov--Kimball model with $w=0.5$ 
at $U=0.3$ (upper panel), $U=0.9$ (middle panel), and $U=1.5$ (lower panel).
Solid lines show mobility edges determined within DMFT with geometric
  averaging and
dashed lines present band edges determined within DMFT with arithmetic averaging.}
\label{fig1}
\end{figure}

Depending on the spectral properties of the Anderson--Falicov--Kimball model
we distinguish three different 
regimes: i) weak interaction regime
for $0<U<W/2$, ii) intermediate interaction regime for $W/2<U\lesssim 1.36W$, and iii)
strong interaction regime for $U\gtrsim 1.36W$. 
Examples of the spectral phase diagrams on the energy--disorder ($\omega-\Delta$) planes in these three
regimes are shown in 
Fig.~\ref{fig1} in the upper, middle, and lower panels, respectively.

\begin{figure}[twb]
\centerline{\includegraphics [clip,width=8.5cm,angle=-0]{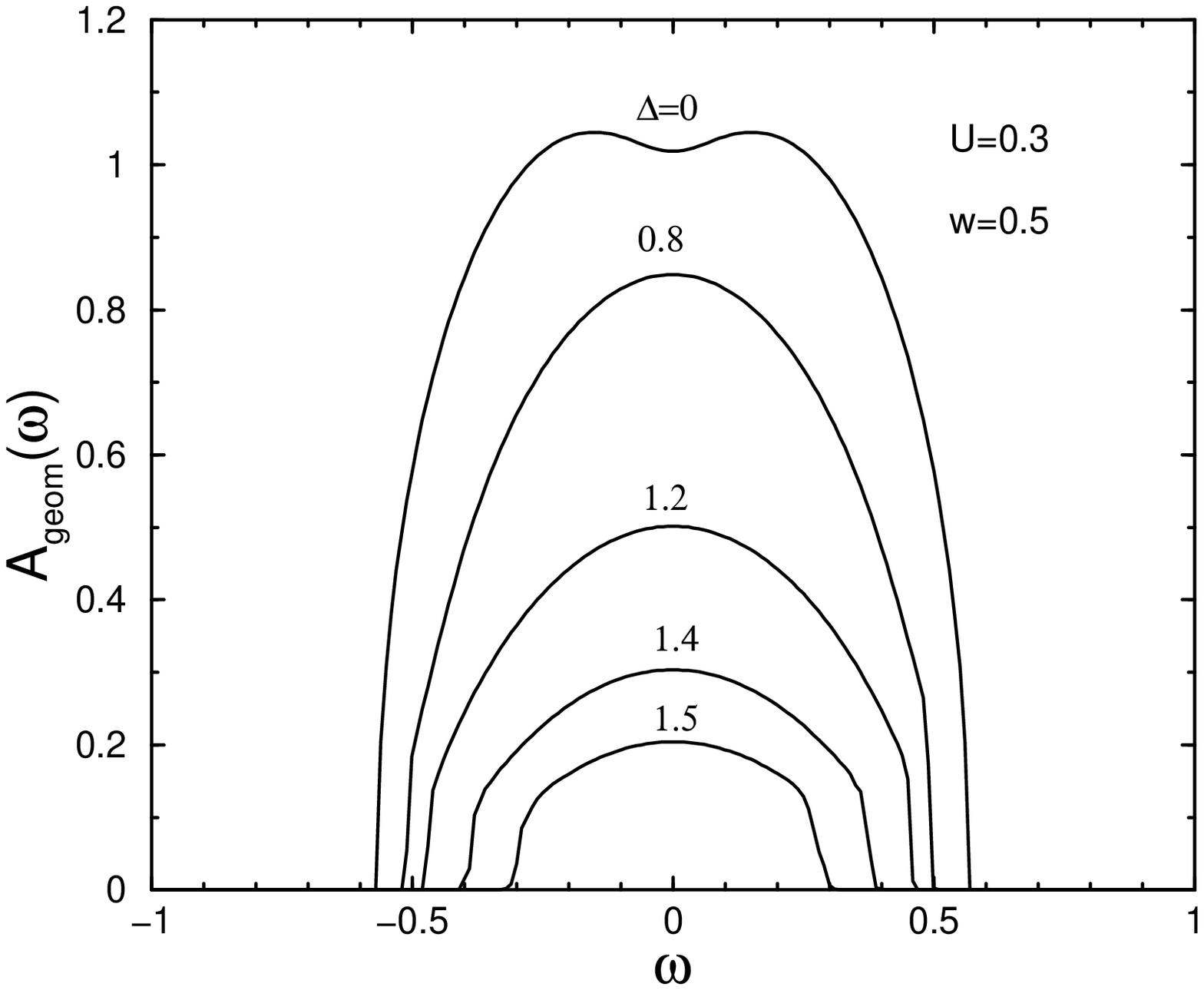}}
\nopagebreak
\centerline{\includegraphics [clip,width=8.5cm,angle=-0]{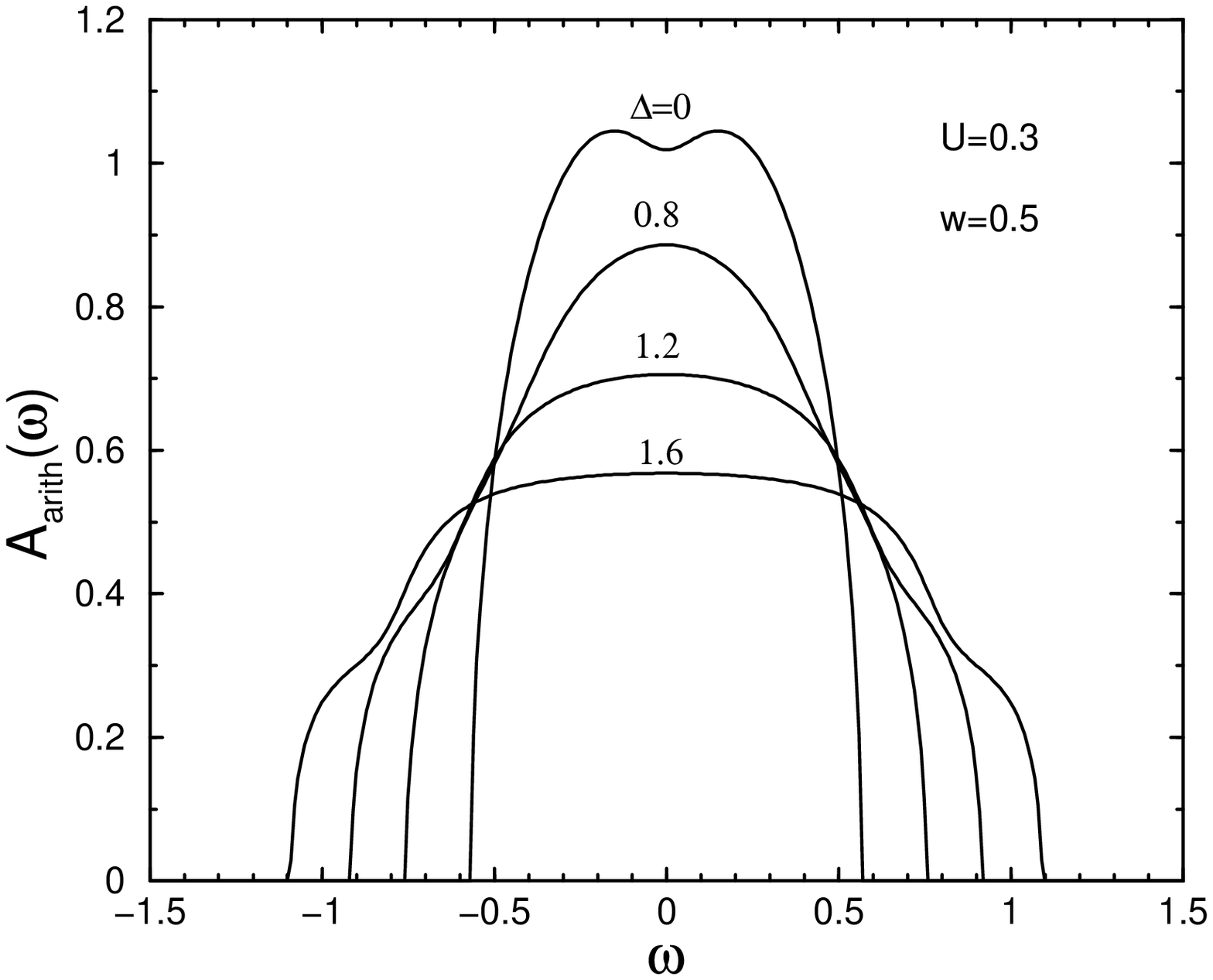}}
\nopagebreak
\caption{Geometrically (upper panel) and arithmetically (lower panel) averaged
  local density of states at $w=0.5$ and 
$U=0.3$  for different disorder strength $\Delta$.
Vanishing of $A_{\rm geom}$ upon increasing $\Delta$ is
a signature of the Anderson localization.}
\label{fig2a}
\end{figure}

In the weak interaction regime (i) the Mott gap is not opened. Increasing the
disorder strength $\Delta$ leads to narrowing of the spectrum with extended
gapless states (continuous 
spectrum) and to broadening of the total band--width.
This is illustrated in  Fig.~\ref{fig2a}
presenting the evolution of $A_{\rm geom}(\omega)$ and $A_{\rm
  arith}(\omega)$, upper and lower panels respectively, upon increasing $\Delta$.   
The continuous spectrum corresponds to a support of $A_{\rm geom}(\omega)$,
i.e. such an energy window for which $A_{\rm geom}(\omega)>0$,
whereas the full band is given by the support of $A_{\rm
  arith}(\omega)$. 
The trajectories of the band edges (dashed lines) determined within
the DMFT with arithmetic averaging and the trajectory of mobility edge (solid
lines) determined within the DMFT with geometric averaging are shown in
Fig.~\ref{fig1}. In the weak interaction regime the localized gapless states
(pure point spectrum) are in a compact part of the spectral 
phase diagram between the mobility and band edges. 
The spectral phase diagram of the Anderson--Falicov-Kimball model in the
weak interaction regime (i) is qualitatively similar to that of the Anderson
model without the interaction.\cite{Dobrosavljevic03}

\begin{figure}[twb]
\centerline{\includegraphics [clip,width=8.5cm,angle=-0]{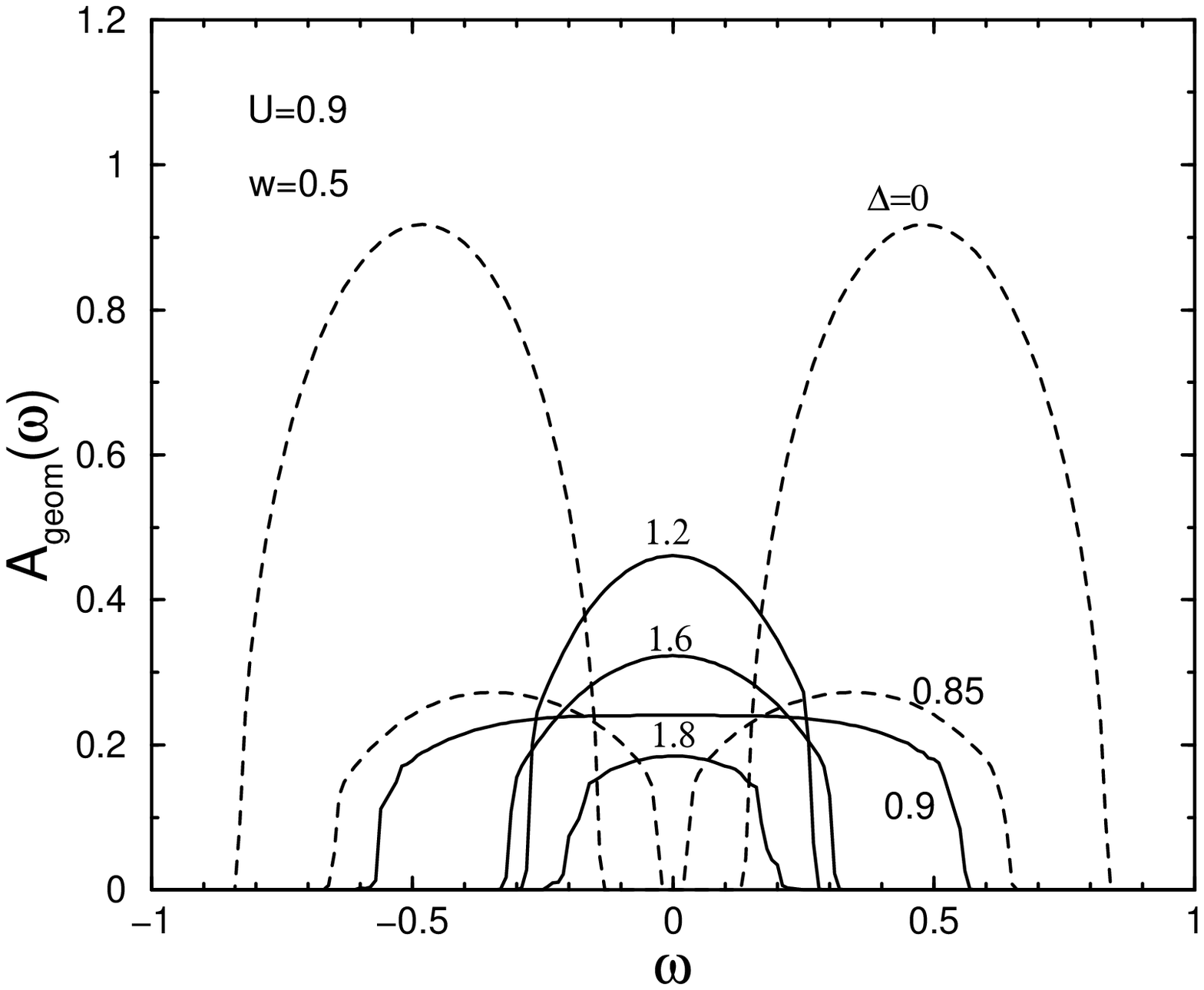}}
\nopagebreak
\centerline{\includegraphics [clip,width=8.5cm,angle=-0]{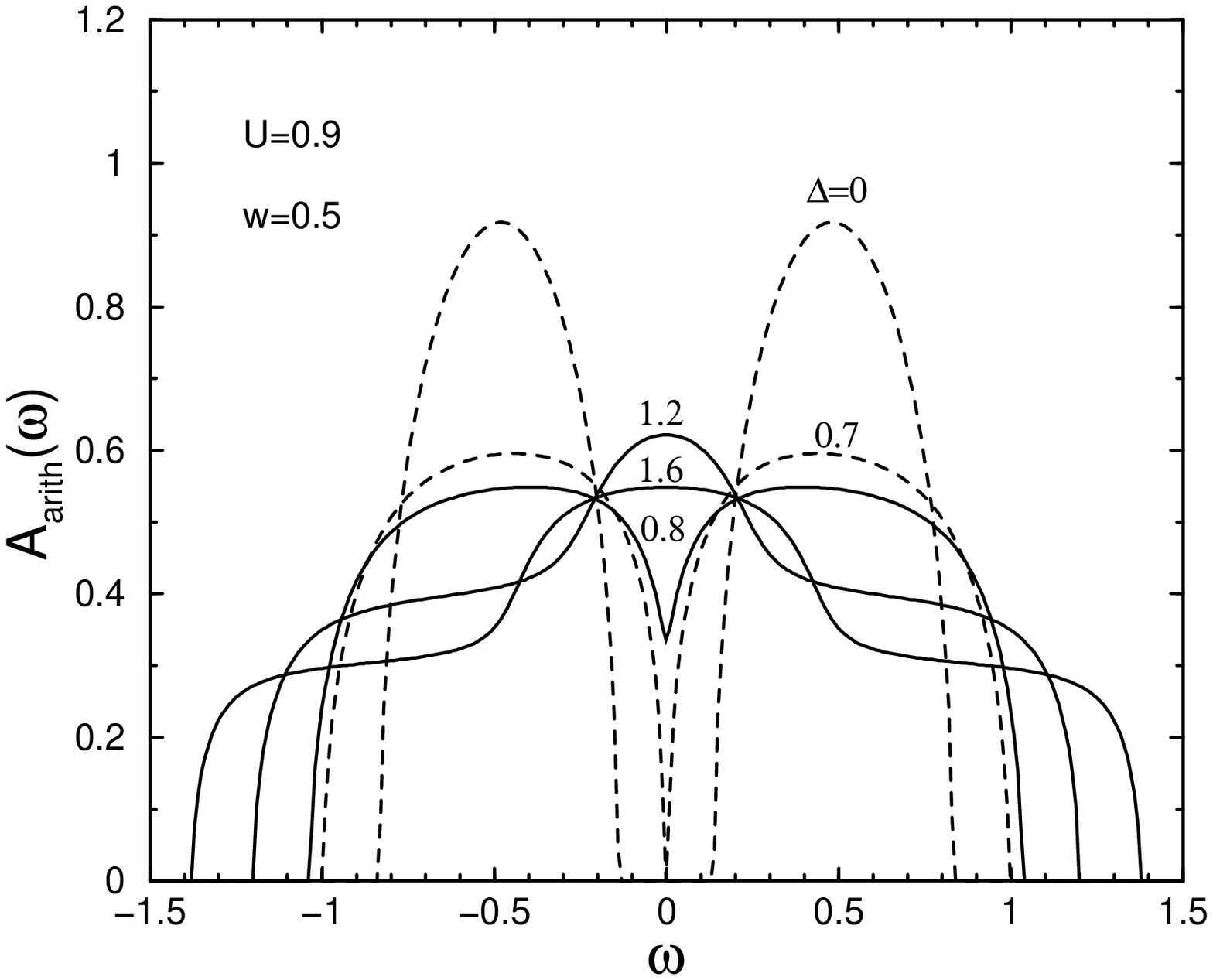}}
\nopagebreak
\caption{The same as in Fig.~\ref{fig2a} but at $U=0.9$.
The correlation (Mott) gap is closed by strong
disorder and the system becomes a bad metal. }
\label{fig2b}
\end{figure}

\begin{figure}[twb]
\centerline{\includegraphics [clip,width=8.5cm,angle=-0]{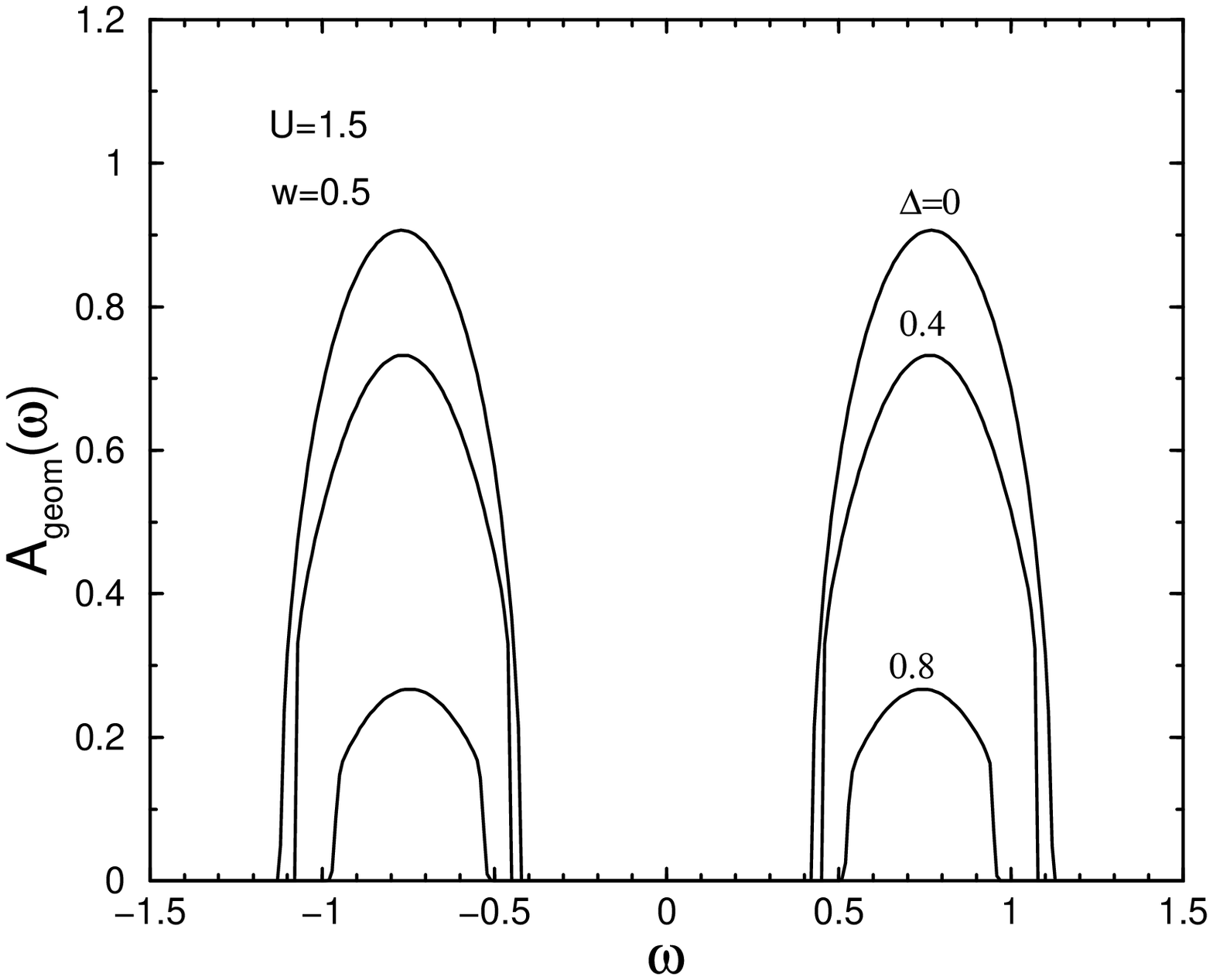}}
\caption{Geometrically averaged local density of states at  $w=0.5$ and
$U=1.5$ for different disorder strength $\Delta$.}
\label{fig2c}
\end{figure}

In the intermediate interaction regime (ii) the Mott gap is opened at
$\Delta=0$, as is shown in  Fig.~\ref{fig2b}. Upon
increasing $\Delta$ the Mott gap is shrunk and finally closed, cf. Fig.~\ref{fig2b}. 
The disorder redistribute the spectral weight filling in the correlation gap
completely by continuous spectrum. The spectral phase diagram is shown in the
middle panel of Fig.~\ref{fig1}. As previously the total band-width,
determined by $A_{\rm  arith}(\omega)$, increases by increasing $\Delta$. 
However in contrast to  (i), in the present case there are two
trajectories  representing external (from the band gap sides)
and internal (from the Mott gap sides) mobility edges. 
Similarly there are two band--edge trajectories, external and
internal ones. 
The extended gapless states are bounded between mobility edge
trajectories as is shown in the middle panel of Fig.~\ref{fig1}. Two separated regions
with localized gapless states are bounded by the mobility edge and the band
edge trajectories. 
If the correlation gap is opened, the localized states
appear in the spectrum from each side of the Hubbard subbands. 

In the strong interaction regime (iii) the Mott gap, determined within the
DMFT framework with geometrical averaging, is never filled in even
at large disorder.  The
LDOS given by $A_{\rm  geom}(\omega)$ and shown in 
 Fig.~\ref{fig2c} always has two separate parts, remnants of the
lower and  the upper Hubbard subbands. This is in contrast with the DMFT
framework with arithmetic averaging where these two subbands always merge if
$\Delta$ is 
sufficiently large. It means that the Mott gap is only filled in by  localized
states when the disorder increases.  
In the spectral phase diagram the spectrum of extended states is given by two
separate lobes bounded by the mobility edge trajectories, as in the lower
panel of Fig.~\ref{fig1}. At  $\Delta>0$ these lobes are surrounded by 
 localized gapless states.

\section{Band center}

\begin{figure}[twb]
\centerline{\includegraphics [clip,width=8.5cm,angle=-0]{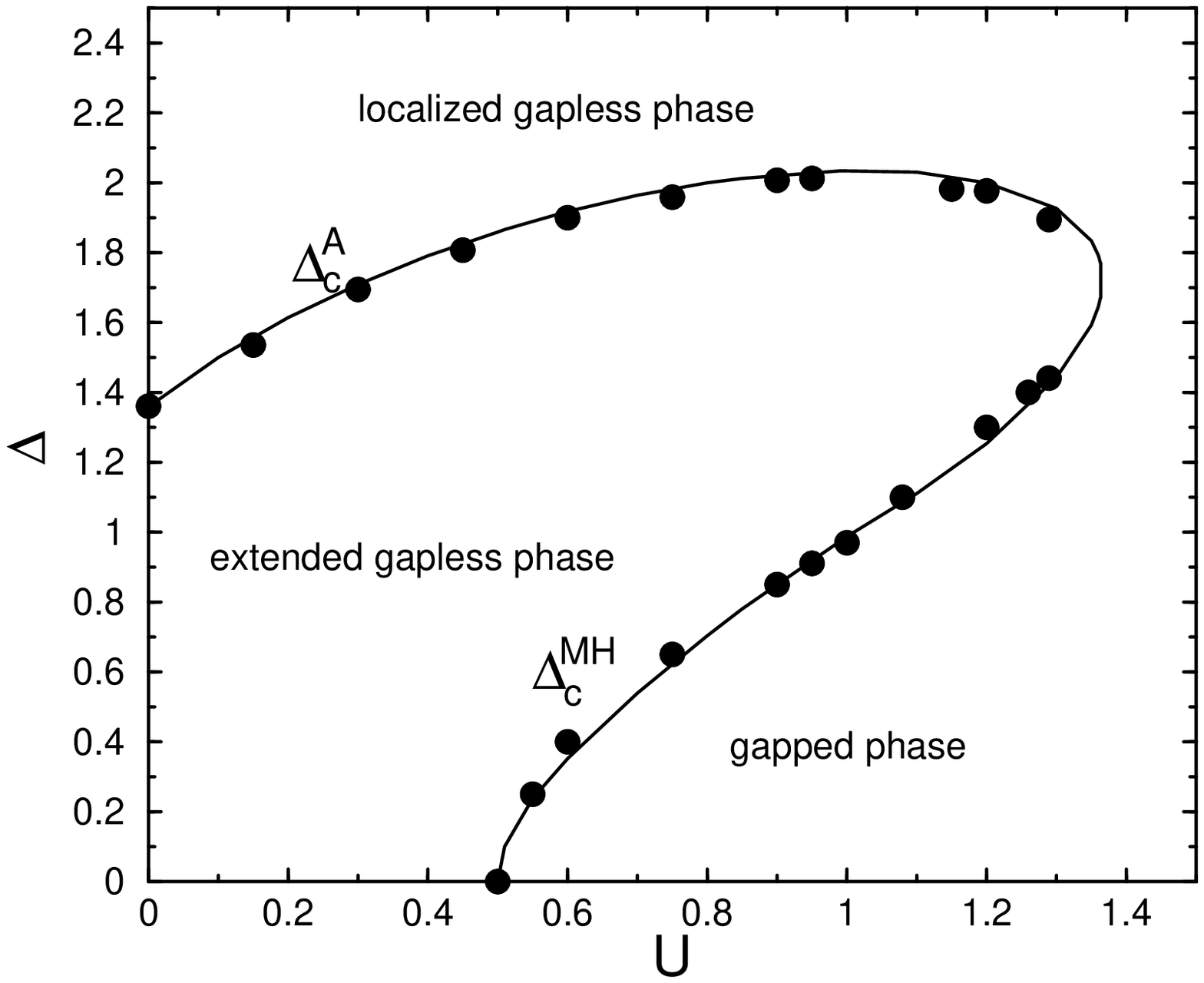}}
\nopagebreak
\centerline{\includegraphics [clip,width=8.5cm,angle=-0]{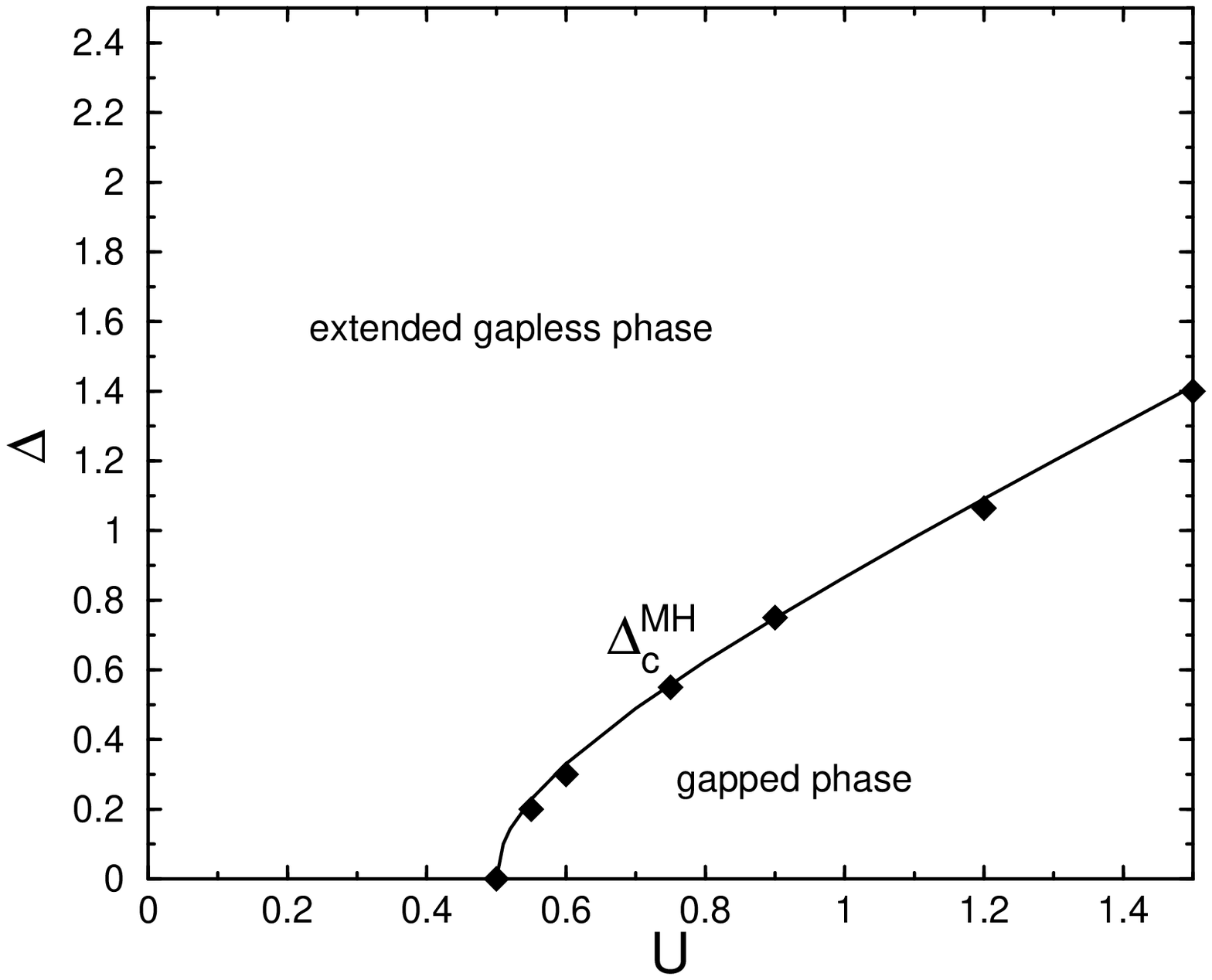}}
\nopagebreak
\caption{Ground state phase diagrams for particles in a band center
determined by using geometric (upper panel) and arithmetic (lower panel)
means.
Dots and squares are determined from the numerical solution of the DMFT
equations. Solid lines are obtained analytically from the linearized DMFT.}
\label{fig4}
\end{figure}

In the half--filled band case the ground state properties are solely
determined by the character of  quantum states in the band center
($\omega=0$). 
Corresponding phase diagrams in the interaction--disorder ($U-\Delta$)  plane are shown 
 in Fig.~\ref{fig4} and discussed below. 
We  define three phases: a) \emph{extended gapless phase} (i.e. a gapless phase with extended
states at the Fermi level), b) \emph{localized gapless phase} 
(i.e. a gapless phase  with localized states at the Fermi level), and c) \emph{gapped phase}
(i.e. a phase with a gap at the Fermi level).

\begin{figure}[twb]
\centerline{\includegraphics [clip,width=8.5cm,angle=-0]{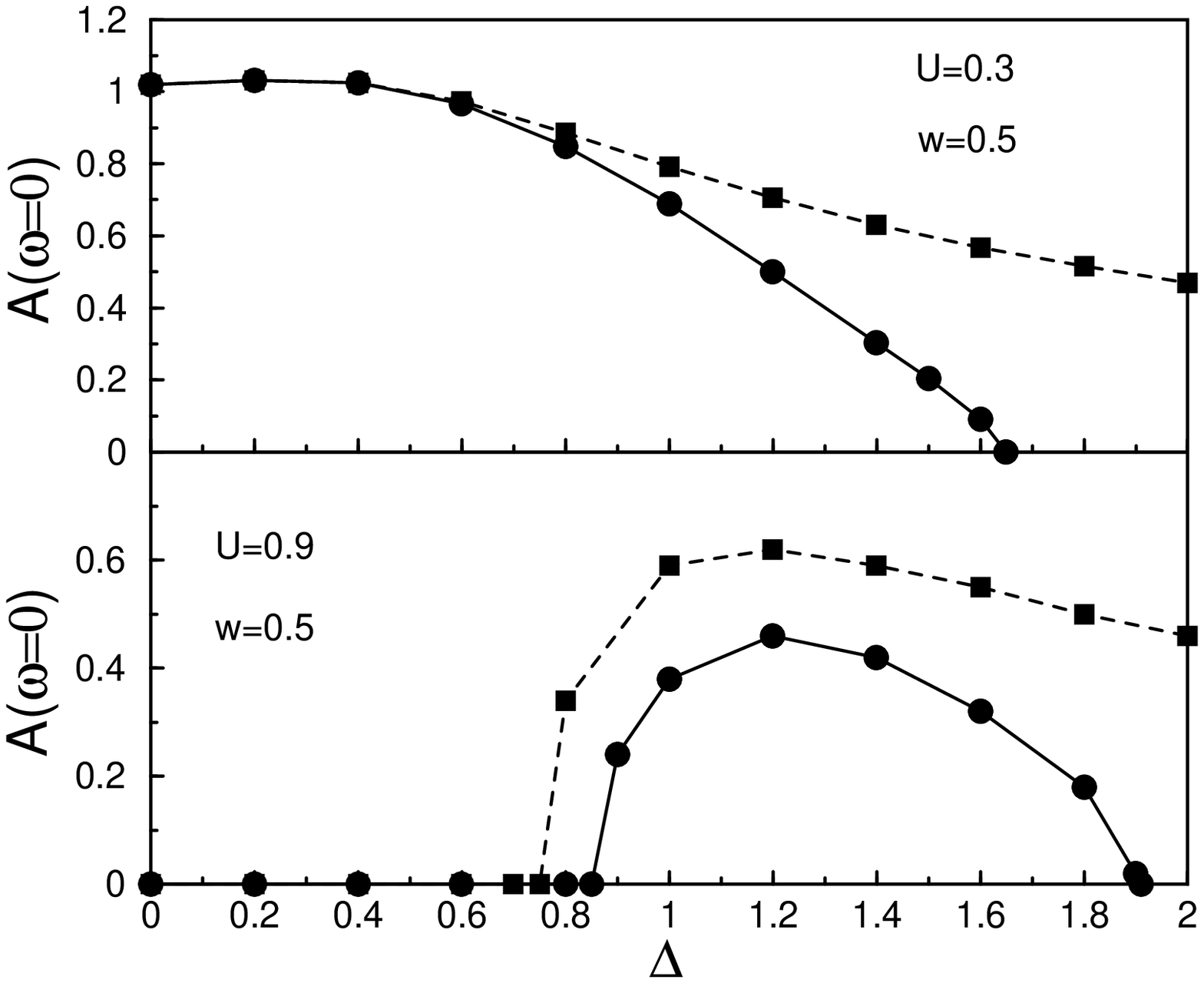}}
\caption{Local density of states in a band center ($\omega=0$) as a function
  of disorder $\Delta$ at $w=0.5$ with $U=0.3$ (upper panel)
and $U=0.9$ (lower panel). Solid (dashed) lines are determined by using geometric (arithmetic)
averaging.}
\label{fig5}
\end{figure}

\begin{figure}[twb]
\centerline{\includegraphics [clip,width=8.5cm,angle=-0]{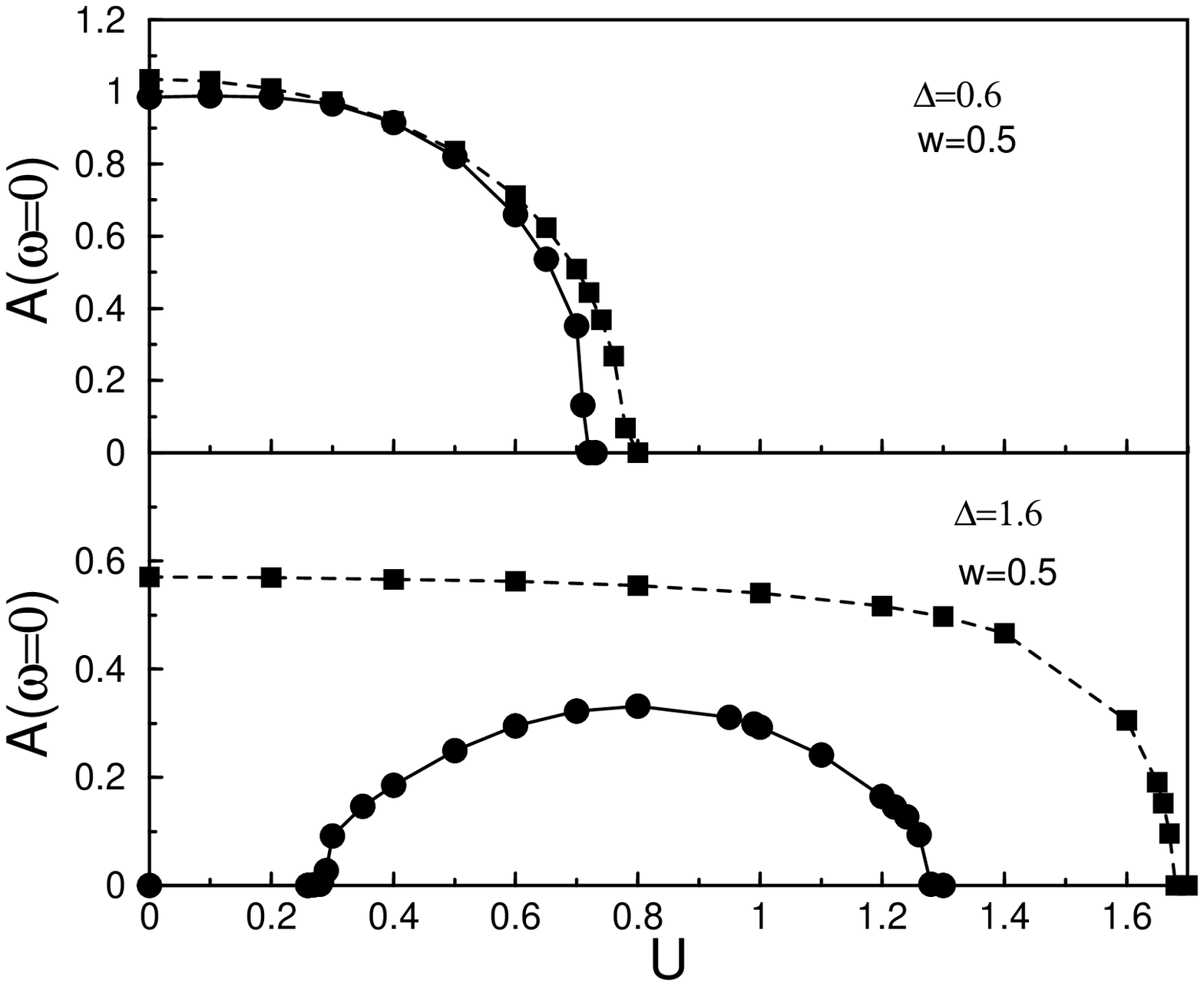}}
\caption{Local density of states in a band center ($\omega=0$)  as a function
  of interaction $U$ at $w=0.5$ with $\Delta=0.6$ (upper panel)
and $\Delta=1.6$ (lower panel). Solid (dashed) lines are determined by using geometric (arithmetic)
averaging.}
\label{fig6}
\end{figure}

The extended gapless phase (disorder metallic phase), is characterized by a
non--zero value of the LDOS. 
In the pure Falicov--Kimball model the Luttinger theorem is not satisfied 
and quasiparticles at  the Fermi level have finite life-time.\cite{si}
It means that due to  the interaction 
 $A_{\alpha}(\omega=0)<N_0(\omega=0)$ even at $\Delta=0$.
Increasing  $\Delta$ at fixed $U$ leads to further decreasing of the LDOS as is
shown in  the upper panel of Fig.~\ref{fig5}.
Similarly, increasing $U$ at constant $\Delta$ leads to  decreasing of the LDOS as 
is presented in the upper panel of Fig.~\ref{fig6}.

Mott-Hubbard MIT, represented by $\Delta^{MH}_c(U)$ lines in Fig.~\ref{fig4},
occurs at small and intermediate disorder $0\leq 
\Delta \lesssim 1.70 W$
and the interaction $W/2\leq U\lesssim 1.36W$. This MIT is continuous  one as
is seen  in the lower panel of Fig.~\ref{fig5} around $\Delta\approx 0.8$
and in Fig.~\ref{fig6} around $U\approx 0.7$ (upper panel) and $U\approx
1.3$ (lower panel). 
 Increasing $\Delta$ above $\Delta_c^{MH}(U)$ when $W/2<U\lesssim 1.36W$ results in a transition from 
a correlated gapped insulator into a bad metallic phase, as 
is illustrated in the lower panel of Fig.~\ref{fig5}. It means that the disorder stabilizes the metallic 
phase. 
Similar Mott--Hubbard MIT is obtained within the DMFT with arithmetic
averaging and the corresponding phase diagram is reproduced in the lower panel
of Fig.~\ref{fig4}.

Anderson MIT line $\Delta_c^A(U)$ is an increasing function of $U$ for 
$0\leq U\lesssim 0.95W$ starting from the value $\Delta_c^A=eW/2\approx
1.36 W$ ($e\approx 2.718$ is the Euler constant)
 in the non--interacting case.\cite{Dobrosavljevic03} 
The interaction impedes the localization of particles due to impurity scattering. 
In particular, for $e W/2\lesssim\Delta\lesssim 2.03W$ and $0<U\lesssim 0.95 W$ the
interaction  turns the Anderson insulator into a bad correlated metal, as is
shown in the lower panel of Fig.~\ref{fig6}.

Mott and Anderson insulators are rigorously defined only for  $U>W/2$ with
$\Delta =0$ and  only for $\Delta >eW/2$ with $U=0$, respectively.
In the presence of the interaction and disorder this distinction can no
longer be made. However, as long as the LDOS shows the
characteristic Hubbard subbands (see Fig.~\ref{fig2c})
one may refer to a \emph{disordered Mott insulator} (gapped phase). With increasing
$\Delta $ the spectral weight of the Hubbard subbands vanishes  and the system
becomes a \emph{correlated Anderson 
insulator} (localized gapless phase). The border between these two types of
insulators occurs  
at $\Delta(U)\approx eW/2\sqrt{2}$ when $U\gg W$. To estimate this value we
used the analytical result from Ref.~\onlinecite{Dobrosavljevic03}.
Our estimation is exact  when $U\to \infty$ since the band-width of the Hubbard satellites 
is rigorously known to be $W/\sqrt{2}$ for the Falicov--Kimball model.\cite{gebhard}
 The results obtained here within DMFT
show that the Mott and Anderson insulators are  continuously connected.
Hence, by changing $U$ and $\Delta$
it is possible to move from one type of the insulator to the other without
crossing the metallic phase.
This is plausible because the Anderson MIT ($U=0$) and the Mott--Hubbard MIT ($\Delta=0$) are
 not associated with a symmetry breaking. 

\section{Linearized dynamical mean--field theory}

At the MIT  (dots and squares in Fig.~\ref{fig4}) the LDOS vanishes in the band center.
Therefore, in the vicinity of the MIT  but on the metallic side 
the LDOS is arbitrary small and 
the transition points on the phase diagram can be determined analytically by linearizing 
the DMFT equations.\cite{Dobrosavljevic03,bullaldmft}
In the band center, due to a symmetry of $A_{\alpha}(\omega)$, we find that 
$G(0)=-i \pi A_{\alpha}(0)$ and is purely imaginary.
Hence the DMFT self-consistency leads to the following recursive relation 
$\eta^{(n+1)}(0)=-i \pi W^2 A^{(n)}_{\alpha}(0)/16$, where the left hand side in the $(n+1)$-th 
iteration step 
is given by the result from $(n)$-th iteration step.
Using Eqs.~\ref{green}~with~\ref{sigma} and expanding them with respect to small $A^{(n)}_{\alpha}(0)$
we find from Eq.~\ref{ldoss} that
\begin{equation}
A^{(n+1)}(0,\epsilon_i)=\frac{W^2}{16}A^{(n)}_{\alpha}(0) \Upsilon(\epsilon_i),
\end{equation}
where
\begin{equation}
\Upsilon(\epsilon)=\frac{\epsilon^2+\left(\frac{U}{2}\right)^2}{
\left[ \epsilon^2-\left(\frac{U}{2}\right)^2\right]^2}.
\end{equation}
The recursive relations within the linearized DMFT (L-DMFT) with geometrical averaging are
\begin{equation}
A_{\rm geom}^{(n+1)}(0)=A_{\rm geom}^{(n)}(0)\frac{W^2}{16}\exp\left[
\frac{1}{\Delta}\int_{-\Delta/2}^{\Delta/2}d\epsilon \ln  \Upsilon(\epsilon)
\right]
,
\end{equation}
and within L-DMFT  with arithmetical averaging are
\begin{equation}
A_{\rm arith}^{(n+1)}(0)=A_{\rm arith}^{(n)}(0)\frac{W^2}{16}\left[
\frac{1}{\Delta}\int_{-\Delta/2}^{\Delta/2}d\epsilon \Upsilon(\epsilon)
\right].
\end{equation}

In a metallic phase the recursions are increasing, i.e. $A_{\alpha}^{(n+1)}(0)>A_{\alpha}^{(n)}(0)$,
whereas in the insulating phase they are decreasing. Therefore, at the boundary curves between metallic 
and insulating solutions in Fig.~\ref{fig4} the recursions are constant 
$A_{\alpha}^{(n+1)}(0)=A_{\alpha}^{(n)}(0)$.
This observation  leads directly to the exact (within DMFT) equations determining the 
curves $\Delta=\Delta(U)$, i.e.
\begin{equation}
1=\frac{W^2}{16}\exp\left[
\frac{1}{\Delta}\int_{-\Delta/2}^{\Delta/2}d\epsilon \ln  \Upsilon(\epsilon)
\right]\equiv \frac{W^2}{16}\exp \left[I_{\rm geom}(U,\Delta)\right],\label{geom}
\end{equation}
for L-DMFT with geometrical averaging, and 
\begin{equation}
1=\frac{W^2}{16}\left[
\frac{1}{\Delta}\int_{-\Delta/2}^{\Delta/2}d\epsilon \Upsilon(\epsilon)
\right]\equiv \frac{W^2}{16} I_{\rm arith}(U,\Delta),\label{arit}
\end{equation}
for L-DMFT with arithmetical averaging.
Both integrals can be evaluated analytically with the results
\begin{eqnarray}
I_{\rm geom}(U,\Delta)=2+\nonumber \\
\ln
\left[ 
\left( \frac{U}{2} \right)^2+
\left( \frac{\Delta}{2}\right)^2 \right]-
2\ln 
\left[ \left( \frac{U}{2} \right)^2-  
\left( \frac{\Delta}{2}\right)^2 \right] + \nonumber \\
\frac{2U}{\Delta}\left[
\arctan \left(\frac{\Delta}{U} \right)-\ln \Big| \frac{\Delta+U}{\Delta-U} \Big|
 \right],
\end{eqnarray}
and 
\begin{equation}
I_{\rm arith}(U,\Delta)=
\frac{1}{
\left( \frac{U}{2} \right)^2-\left( \frac{\Delta}{2}\right)^2
}.
\end{equation}
Solutions of Eqs.~\ref{geom}~and~\ref{arit} are shown as  solid curves  
in the upper and the lower panels of Fig.~\ref{fig4}, respectively.
We find excellent agreement with the numerical solutions 
of the full DMFT equations (dots and squares in Fig.~\ref{fig4}).
At small $U$ the critical disorder strength obtained from (\ref{geom})
 increases linearly with the interaction, i.e. 
$\Delta(U)\approx W e/2+\pi U /2$.
This is  because the total bandwidth increases linearly with $U$.
At small $\Delta$ the solution of Eq.~\ref{geom} (L-DMFT with geometrical averaging) 
is found to be $\Delta(U)\approx \sqrt{U^2-(W/2)^2}$.
It turns out that the curve $\Delta(U)= \sqrt{U^2-(W/2)^2}$ is also an exact solution of 
Eq.~\ref{arit} (L-DMFT with arithmetical averaging) for all $U\geq W/2$. 
At weak disorder both approaches give the same results. 

\section{Conclusions}

In the present paper we introduced the Anderson--Falicov--Kimball model and
solved it obtaining the phase diagrams for the ground state with the
suppressed long--range order. Three different phases, the disordered metal, the
disordered Mott insulator, and the correlated Anderson insulator, were
identified. It was shown that correlation and disorder compete with each other
stabilizing the metallic phase against occurring one of the insulators. We found that
these two insulators are continously connected.

The phase diagram with the three phases in the ground state for the
Anderson--Falicov--Kimball model is similar to the phase diagram for the
Anderson--Hubbard model solved within the DMFT with geometric averaging in
Ref.~\onlinecite{byczuk04}. There are however important qualitative
differences between these two solutions. 
In the Anderson--Falicov--Kimball model the Mott transition is continuous
whereas in the Anderson--Hubbard model there is a hysteresis at low 
and a crossover transition at high disorder strengths. In addition, in the
Anderson--Hubbard model the Luttinger pinning in the disordered metal is
reconstructed by strong correlations. This feature is absent in the
Anderson--Falicov--Kimball model, where the Luttinger pinning is violated even
in the pure case.\cite{si} 

A similar technique, i.e. the DMFT with geometric averaging, could be used to
solve other models with disorder and interaction 
between quantum (mobile) and classical (immobile) degrees of freedom.\cite{lem}
 In such cases
the self--energy should be given analytically and this removes the problem of
using any numerical impurity solver, as was necessary in the
Anderson--Hubbard model.\cite{byczuk04}
Then not only the LDOS at the Fermi level but also
mobility and band edge trajectories can be easily determined. 
Such models are important for understanding the physics of
manganites\cite{dagotto01} or diluted magnetic
semiconductors,\cite{spintronics,dietl02} where charge carriers are 
coupled to randomly distributed localized magnetic moments. The role of
the disorder and the Anderson localization are inherent for those correlated
systems.

\begin{acknowledgments}

It is a pleasure to thank  R.~Lema\'nski for encouraging 
discussion on this project and critical reading of the manuscript
The author also thanks R.~Bulla  and D.~Vollahardt for useful discussion
on different aspects of DMFT and Anderson localization.
Financial support through KBN-2 P03B 08 224 is  acknowledged.
This work was also supported in part by the Sonderforschungsbereich 484 of
the Deutsche Forschungsgemeinschaft. 
\end{acknowledgments}



\begin{references}
\bibitem{definition} The most probable value of a random quantity
is defined as that value for which its probability distribution function  becomes maximal.
\bibitem{comment2} For a uniform probability distribution function  and $N$
  discrete values of $A(\omega,\epsilon_i)$ one finds  
$A_{\rm{geom}}(\omega)=\left(\prod_{i=1}^N 
A(\omega,\epsilon_i)
\right)^{1/N}=
\exp\{[\sum_{i=1}^N\ln A(\omega,\epsilon_i)]/N\}$. Hence $A_{\rm{geom}}(\omega)$ 
vanishes if any of the $A(\omega,\epsilon_i)$ is zero. The arithmetic average does not have such a property.
\end{references}

\end{document}